\documentclass[aps,prl,showpacs,floatfix,twocolumn]{revtex4}
\usepackage{latexsym}
\usepackage{graphicx,subfigure}
\begin{document}
\draft
\thispagestyle{empty}

\title{Is there a black hole minimum mass?}
\author{Tomohiro Harada~\footnote{Electronic
address:harada@rikkyo.ac.jp}}
\affiliation{
Department of Physics, Rikkyo University, Toshima, 
Tokyo 171-8501, Japan}

\date{\today}
\begin{abstract}
Applying the first and generalised second laws of thermodynamics for a
realistic process of near critical black hole formation, 
we derive an entropy bound, which is identical to 
Bekenstein's one for radiation.
Relying upon this bound,
we derive an absolute minimum mass $\sim0.04 \sqrt{g_{*}}m_{\rm Pl}$, 
where $g_{*}$ and $m_{\rm Pl}$ is the effective degrees of 
freedom for the initial temparature and 
the Planck mass, respectively.
Since this minimum mass coincides with the lower bound on
masses of which black holes can be regarded as classical against
the Hawking evaporation, the thermodynamical argument 
will not prohibit the formation of the 
smallest classical black hole.
For more general situations, we derive a minimum mass,
which may depend on the initial value for entropy per particle.
For primordial black holes, however, 
we show that this minimum mass can not be much greater than 
the Planck mass at any formation epoch of the Universe, as long as
$g_{*}$ is within a reasonable range.
We also derive a size-independent upper bound on 
the entropy density of a stiff fluid 
in terms of the energy density.
\end{abstract}
\pacs{04.70.Dy, 04.70.Bw, 04.20.Dw}
\maketitle

Choptuik~\cite{choptuik1993} numerically studied a self-gravitating system of 
a massless scalar field and revealed that there appear
critical phenomena at the threshold of black hole formation.
If a generic one-parameter ($\lambda$) family of initial data sets are prepared, 
the black hole mass $M$ resulting from the time evolution 
obeys a scaling law $M\propto |\lambda-\lambda^{*}|^{\gamma}$ for 
$\lambda\approx \lambda^{*}$, where $\lambda^{*}$
is the critical value and $\gamma (>0)$ is the critical exponent.
The consequence is that there are black holes of 
arbitrarily small masses in the limit $\lambda \to \lambda^{*}$.
This work has been generalised to a system of 
a radiation fluid~\cite{ec1994}.
In the cosmological context, critical phenomena at the threshold 
of primordial black hole formation have been found~\cite{nj1999}
but subsequently Hawke and Stewart~\cite{hs2002} reported that there is 
a minimum mass $\sim 2\times 10^{-4}$ times the mass $M_{\rm H}$
contained within the Hubble horizon due to shock wave formation based on 
numerical simulations with high-resolution shock capturing scheme
and suggested the link with kink instability~\cite{harada1998}.
Here we discuss another possibility of minimum masses
based on black hole thermodynamics.
Complementarily, the maximum masses of primordial black holes
for different formation scenarios have been studied~\cite{carr1975,hc2005}.
We adopt the units in which $c=\hbar=k=1$ and denote $G=m_{\rm Pl}^{-2}$.

The final state of complete gravitational collapse 
is an outgoing flux and a Kerr black hole, for which 
the area $A$ of the event horizon 
is given by $A=4\pi m_{\rm Pl}^{-4}\{[M+(M^{2}-
(J/M)^{2})^{1/2}]^{2}+(J/M)^{2}\}$,
where $J$ is the angular momentum of the hole.
Then, neglecting the rotation, the 
Bekenstein-Hawking entropy $S_{\rm BH}$ is given by 
\begin{equation}
S_{\rm BH}=\frac{m_{\rm Pl}^{2}}{4}A
\simeq 4\pi \frac{M^{2}}{m_{\rm Pl}^{2}}.
\end{equation}
The effects of rotation may affect the discussion 
by a factor of two or so.

Chisholm~\cite{chisholm2006v1} implicitly assumes that the region which 
is going to be a black hole is isolated from its environment, 
applies the generalised second law of 
thermodynamics for the collapse of a radiation fluid, and 
gets a lower bound on (primordial) black hole masses as $M\gtrsim 
m_{\rm Pl}^{2}/T_{\rm i}\sim 
g_{*}^{1/4}m_{\rm Pl}^{2}\epsilon_{\rm i}^{-1/4}$
and a lower bound on the mass fraction $f$ of the 
primordial black hole $M$ to the horizon mass $M_{\rm H}$ 
as $f\gtrsim \sqrt{g_{*}} T_{\rm i}/m_{\rm Pl}
=(g_{*}\epsilon_{\rm i})^{1/4}/m_{\rm Pl}$,
where $g_{*}$, $T$ and $\epsilon$ are 
the effective degrees of freedom of relativistic particles,
the temperature and the energy density, respectively,
and the suffix ``i'' indicates it is for the initial value of the collapse.

Here we adopt more realistic assumption that 
the region to be a black hole is surrounded by the outer region 
with nonnegligible pressure $p$ rather than isolated.
This assumption will be appropriate not only for primordial 
black holes but also for near critical black hole formation 
in an asymptotically flat spacetime.
This external pressure can exert a considerable amount 
of positive work on this region.
Let $R$, $E$ and $S$ 
be the radius, energy and entropy of this region.
Let us assume that self-gravity of this region is sufficiently 
weak at the initial moment for near critical collapse. 
The first law then becomes $d(\epsilon R^{3})=-p d(R^{3})$ for the
adiabatic process, where $p$ is the pressure. 
For a radiation fluid $p=\epsilon/3$, 
it follows that $\epsilon$ is proportional to $R^{-4}$ and
$E$ is proportional to $R^{-1}$.
This energy increase is due to the work term exerted by the 
external pressure.
When the radius of the region is approximately equal to the 
gravitational radius, i.e., 
$M\simeq (R_{\rm i}/R_{\rm g})E_{\rm i}$, where 
$R_{\rm g}\simeq 2 M/m_{\rm Pl}^{2}$, a black hole forms.
This yields 
\begin{equation}
M^{2}\simeq \frac{m_{\rm Pl}^{2}}{2} E_{\rm i} R_{\rm i}.
\end{equation}
Applying the generalised second law $S_{\rm BH}\ge S_{\rm i}$, we get 
\begin{equation}
 2\pi E_{\rm i} R_{\rm i}\ge S_{\rm i}.
\label{eq:entropy_bound_rad}
\end{equation}
This is identical to the entropy bound that Bekenstein~\cite{bekenstein1981}
invented from a different derivation.
It should be emphasised that we can only have a trivial lower bound
\begin{equation}
 M\geq m_{\rm Pl}\sqrt{\frac{S_{\rm i}}{4\pi}}
\label{eq:minimum_mass_entropy}
\end{equation}
on black hole masses at this stage unlike 
Chisholm~\cite{chisholm2006v1}'s argument 
based on the assumption of isolation. 

Below we see that the right-hand side of 
Eq.~(\ref{eq:minimum_mass_entropy}) is also bounded from below
through the entropy bound.
Equation~(\ref{eq:entropy_bound_rad}) 
implies that the radius of the region destined to be 
a black hole must be larger than $S/(2\pi E)$. Assuming that 
the region to be a black hole is approximately homogeneous,
the radius of this region is bounded from below as
\begin{eqnarray}
R_{\rm i}\gtrsim \frac{2}{3\pi T_{\rm i}}, 
\label{eq:size_bounded_temperature}
\end{eqnarray}
where we have used $s=4\epsilon/(3T)$.
Equation~(\ref{eq:size_bounded_temperature}) might be seen
along the line of Heisenberg's uncertainty relation,
since $T$ gives the characteristic momentum of photons.
The entropy contained in this region is then bounded from below as
\begin{equation}
S_{\rm i}\gtrsim \frac{4\pi}{3}s_{\rm i}\left(\frac{s_{\rm i}}
{2\pi \epsilon_{\rm i}}\right)^{3} =\frac{64}{3645}g_{*},
\end{equation}
where we have used $\epsilon=(\pi^{2}/30)g_{*}T^{4}$.
Then, the generalised second law yields a minimum mass
\begin{equation}
M \gtrsim \frac{4}{27}\sqrt{\frac{g_{*}}{5\pi}}m_{\rm Pl},
\label{eq:minimum_pbh_rad}
\end{equation}
and therefore a lower bound
\begin{equation}
f\gtrsim \frac{16\pi}{405}g_{*}\left(\frac{T}{m_{\rm Pl}}\right)^{2}.
\end{equation}
It should be noted that the above lower bound is very 
different from Chisholm~\cite{chisholm2006v1}'s estimates.
If $g_{*}$ is order 1000, this minimum mass is 
order the Planck mass.
This argument implies that the Planck mass naturally arises 
as a black hole minimum mass even without explicit form of quantum
gravity. This also implies that the generalised second law 
naturally requires quantum gravity and the former should be 
justified by the latter. 
If $g_{*}$ is much larger than order 1000, such as in Hagedorn-type
scenarios~\cite{hagedorn}, the minimum mass 
may become much larger than the Planck mass.

In the process of gravitational collapse to a black hole,
there is a possibility that 
a heat flux may be induced by radiation transfer and/or shocks.
In such a case, the system is no longer adiabatic and cannot 
be described by a perfect fluid alone.
If the heat flux carries a considerable amount of 
heat away from the region within the time scale of collapse,
the entropy in the collapsing region could decrease and 
the resultant black hole mass could be much smaller than the 
estimated value here. 
It will be interesting to investigate this possibility further 
based on specific processes of heat flux generation.

A black hole emits black body radiation due to quantum 
particle creation and its temperature is given 
by $T_{\rm H}=m_{\rm Pl}^{2}/(8\pi M)$~\cite{hawking1974,hawking1975}.
The energy loss rate due to the Hawking radiation for 
a black hole of mass $M$ is estimated as~\cite{bd1982}
\begin{equation}
\frac{dM}{dt}=-\frac{g_{*}\Gamma}{15360\pi}\frac{m_{\rm Pl}^{4}}{M^{2}},
\end{equation}
where $\Gamma$ is a positive constant of order unity. From this, 
the evaporation time $t_{\rm ev}$ is calculated as
\begin{equation}
t_{\rm ev}=\frac{5120\pi}{g_{*}\Gamma}
\left(\frac{M}{m_{\rm Pl}}\right)^{3}m_{\rm Pl}^{-1}.
\end{equation}
Since the classical dynamical time of the black hole 
is estimated by the light crossing time of the horizon radius
and calculated as $t_{\rm dyn}=2(M/m_{\rm Pl})m_{\rm Pl}^{-1}$,
the black hole can be regarded as classical if and only if 
$t_{\rm ev}\gtrsim t_{\rm dyn}$ or 
\begin{equation}
M\gtrsim  \frac{1}{16}\sqrt{\frac{g_{*}\Gamma}{10\pi}}m_{\rm Pl}.
\end{equation}
The right-hand side gives another lower bound for classical 
black holes and this approximately agrees with the minimum 
mass given by Eq.~(\ref{eq:minimum_pbh_rad}) 
from the thermodynamical second law argument up to 
the factor of $\sqrt{g_{*}}$.
In other words, the thermodynamical minimum mass argument 
will not prohibit the formation of the smallest classical 
black holes.

The thermodynamical minimum mass argument can 
be easily extended to a general perfect
fluid with the equation of state $p=\alpha \epsilon$, where $\alpha$
is a constant.
We assume $0\leq \alpha \leq 1$ to respect causality and 
thermodynamical stability. 
The first law then yields $M\simeq 
(R_{\rm i}/R_{g})^{3\alpha} E_{\rm i}$ and we have
\begin{equation}
M=\left(\frac{m_{\rm Pl}^{2}R_{\rm i}}{2}\right)^{3\alpha/(1+3\alpha)}E_{\rm i}^{1/(1+3\alpha)}.
\end{equation}
Applying the generalised second law $S_{\rm BH}\ge S_{\rm i}$, 
we get
\begin{equation}
4\pi m_{\rm Pl}^{-2(1-3\alpha)/(1+3\alpha)}
\left(\frac{R_{\rm i}}{2}\right)^{6\alpha/(1+3\alpha)}
E_{\rm i}^{2/(1+3\alpha)}\ge S_{\rm i}.
\label{eq:general_entropy_bound}
\end{equation}
This is an entropy bound for the general case.
For $0\leq \alpha <1$, this bound can be regarded as providing the 
minimum radius of the region to be a black hole:
\begin{equation}
R_{\rm i}\gtrsim m_{\rm Pl}^{2(1-3\alpha)/[3(1-\alpha)]}
\left(\frac{s_{\rm i}^{1+3\alpha}}{\epsilon_{\rm i}^{2}}
\right)^{1/[3(1-\alpha)]},
\end{equation} 
where and hereafter we omit numerical factors.
The entropy of the collapsing region is bounded from below as
\begin{equation}
S_{\rm i}\simeq 
\frac{4\pi}{3}s_{\rm i}R^{3}_{\rm i}\gtrsim 
m_{\rm Pl}^{2(1-3\alpha)/(1-\alpha)}
\left(\frac{s_{\rm i}}
{\epsilon_{\rm i}^{1/(1+\alpha)}}\right)^{2(1+\alpha)/(1-\alpha)}.
\label{eq:entropy_lower_bound}
\end{equation}
Then, the generalised second law yields
\begin{equation}
M\gtrsim m_{\rm Pl}^{(1-3\alpha)/(1-\alpha)}
\left(\frac{s_{\rm i}}{\epsilon_{\rm i}^{1/(1+\alpha)}}
\right)^{(1+\alpha)/(1-\alpha)}\cdot m_{\rm Pl}.
\label{eq:general_lower_bound}
\end{equation}
Since $\epsilon\propto V^{-(1+\alpha)}$, where $V$ is the volume, 
\begin{equation}
n\equiv g_{*}[\epsilon/(g_{*}m_{\rm Pl}^{4})]^{1/(1+\alpha)} m_{\rm Pl}^{3},
\end{equation}
has the dimension of energy cubed, is proportional to the degrees
of freedom $g_{*}$ and $N\equiv nV$ is constant and nondimensional.
Therefore, we can identify $n$ with the conserved number density 
of particles. Using this fact, we can rewrite 
Eqs.~(\ref{eq:entropy_lower_bound}) and (\ref{eq:general_lower_bound}),
respectively, as
\begin{eqnarray}
S_{\rm i}&\gtrsim& g_{*}^{2\alpha/(1-\alpha)}
\left(\frac{S}{N}\right)^{2(1+\alpha)/(1-\alpha)}, \\
M&\gtrsim& g_{*}^{\alpha/(1-\alpha)}
\left(\frac{S}{N}\right)^{(1+\alpha)/(1-\alpha)}\cdot m_{\rm Pl},
\end{eqnarray}
where $N$ is the conserved number of particles, such as photons.
This means that, in general, the minimum mass can depend
on the entropy per particle.
If $\epsilon=\epsilon(T)$, 
$dE=TdS-pdV$ and the energy equation
$(\partial E/\partial V)_{T}=T(\partial p/\partial T)_{V}-p$
yield $\epsilon=\tilde{\beta} T^{(1+\alpha)/\alpha}$ and 
$s=(1+\alpha)\tilde{\beta}T^{1/\alpha}$,
where $\tilde{\beta}$ is a constant.
We can write $\tilde{\beta}=\beta C^{(3\alpha-1)/\alpha}$, where 
$\beta$ is a nondimensional constant and $C$ is a constant which
has the dimension of energy.
Equations~(\ref{eq:entropy_lower_bound}) and 
(\ref{eq:general_lower_bound}) then become respectively
\begin{eqnarray}
S_{\rm i}&\gtrsim& \beta^{2\alpha/(1-\alpha)} (1+\alpha)^{2(1+\alpha)/(1-\alpha)} 
\left(\frac{m_{\rm Pl}}{C}\right)
^{2(1-3\alpha)/(1-\alpha)}, \\
M&\gtrsim& \beta^{\alpha/(1-\alpha)} (1+\alpha)^{(1+\alpha)/(1-\alpha)} 
\left(\frac{m_{\rm Pl}}{C}\right)
^{(1-3\alpha)/(1-\alpha)}m_{\rm Pl}.
\label{eq:thermal_lower_bound}
\end{eqnarray}
Therefore, if $C \ll m_{\rm Pl}$ for $0<\alpha<1/3$
or $C \gg m_{\rm Pl}$ for $1/3<\alpha<1$, 
the black hole minimum mass gets much larger than the Planck mass.

It should be noted that we need a special treatment for a stiff fluid
$\alpha=1$. A massless scalar field can be also regarded as a stiff
fluid if and only if its gradient is timelike.
In this case, Eq.~(\ref{eq:general_entropy_bound}) does not 
yield a lower bound on $R_{\rm i}$. Instead, we have
a size-independent bound on the entropy density 
in terms of the energy density as
\begin{equation}
s\lesssim m_{\rm Pl}\sqrt{\epsilon}.
\end{equation}
For this case, we have no black hole minimum mass
because there is no lower bound on the entropy contained within the 
region to be a black hole.

For an interesting application, 
let us consider primordial black hole formation.
Because of the second law, 
the ratio $S/N$ is a nondecreasing function of time. 
We can assume that matter fields 
move together due to strong coupling.
In this case, $\epsilon$ and $\alpha$ in 
the above discussion can be regarded
as those for the dominant component in energy.
Since we can neglect the effects of co-existing epochs, this ratio 
before the radiation-dominated era is equal to or 
smaller than the value at the radiation-dominated era, which is order unity.
This means that the minimum mass before the radiation-dominated era
is not much greater than $m_{\rm Pl}$ as long as $g_{*}$ is within
a reasonable range.

After the matter-radiation equality, nonrelativistic matter fields,
such as cold dark matter and baryonic component,
begins to dominate the Universe in energy but most of the entropy 
is still held in background thermal radiation. 
Gravitational collapse in this regime proceeds as the collapse of 
dark matter and the concentration 
of radiation due to the gravitational field will be negligible.
This means that we can treat this process of black hole formation as 
that with $\alpha=0$ and the ratio $s/\epsilon$ becomes effectively 
smaller than order unity because of the escape of entropy from 
the collapsing region. This means that the minimum mass of 
black holes becomes smaller than the Planck mass, although
the application of this argument to black holes much smaller than
the Planck mass is limited.

These considerations lead to the conclusion that since the total 
entropy contained within the Universe does not decrease in time,
which follows from the second law of thermodynamics,
the thermodynamic minimum mass of primordial black holes is not
much greater than $m_{\rm Pl}$ at any formation 
epoch of the Universe, unless $g_{*}$ is extremely large. 
It is expected that the modification of observational 
constraints due to critical behaviour~\cite{yokoyama1998,nj1998,gl1999} 
will not be so sensitive to the existence of minimum mass discussed here.

After this paper was received, Ref.~\cite{chisholm2006v1} has been 
significantly revised and published as Ref.~\cite{chisholm2006}.
The lower bound obtained in~\cite{chisholm2006} is very similar to
Eq.~(\ref{eq:minimum_pbh_rad}) in the current paper.
However, they are very different in derivation and therefore assumptions. 
In Ref.~\cite{chisholm2006}, the lower bound directly comes from the 
entropy of radiation immediately before the black hole formation.
Hence, $g_{*}$ is regarded as $g_{*}(T_{\rm f})$ 
for the temperature $T_{\rm f}$ at the time of formation.
The discussion there crucially relies upon 
the assumption that the matter field 
is a radiation fluid at the formation.
Moreover, since self-gravity is of course very strong 
and the system is highly dynamical
immediately before the horizon formation,
the assumption that entropy only comes from 
radiation yet needs to be checked. In fact, 
Penrose~\cite{penrose1979} introduced 
the concept that entropy can reside in a
strong gravitational field and this is called gravitational entropy. 
See also~\cite{mena_tavakol1999}
and references therein. The contribution from gravitational entropy 
could in principle significantly change the total entropy 
immediately before the black hole formation.
In the current paper, in contrast, since the derivation is given in terms of 
the initial configuration well before the black hole formation, 
$g_{*}$ can be regarded as $g_{*}(T_{\rm i})$ for the initial temperature
$T_{\rm i}$ and the assumption of weak self-gravity is justifiable
in a self-consistent manner.
Even if we take uncertainty in matter field models 
at very high energy scale and/or strong self-gravity at horizon
formation into account, the thermodynamical 
argument developed in this paper will be modified only 
in terms of small correction terms.

\acknowledgments
The author would like to thank K.~Nakamura, H.~Maeda, J.~Yokoyama 
and S.~Mukohyama for 
discussions and helpful comments.
He would also thank the anonymous referee for 
her or his helpful comments. 
This work was partly supported by the Grant-in-Aid for 
Young Scientists (B), 18740144, from 
the Ministry of Education, Culture, Sports,
Science and Technology (MEXT) of Japan.


\begin{thebibliography}{99}
\bibitem{choptuik1993}
M.~W.~Choptuik, 
Phys. Rev. Lett. {\bf 70}, 9 (1993).
\bibitem{ec1994}
C. R. Evans and J. S. Coleman, 
Phys. Rev. Lett. {\bf 72}, 1782 (1994). 
\bibitem{nj1999}
J.~C.~Niemeyer and K.~Jedamzik,
Phys. Rev. D{\bf 59}, 124013 (1999). 
\bibitem{hs2002}
I.~Hawke and J.~M.~Stewart, Class. Quantum. Grav. {\bf 19}, 3687 (2002).
\bibitem{harada1998}
T.~Harada, 
Class. Quantum Grav. {\bf 18}, 4549 (2001).
\bibitem{carr1975}
B.~J.~Carr, Astrophys. J. {\bf 201}, 1 (1975).
\bibitem{hc2005}
T.~Harada and B.~J.~Carr, Phys. Rev. D{\bf 71}, 104009 (2005).
\bibitem{chisholm2006v1}
J. R. Chisholm, astro-ph/0604174 v1.
\bibitem{bekenstein1981}
J.~D.~Bekenstein, Phys. Rev. D{\bf 23}, 287 (1981).
\bibitem{hagedorn}
R.~Hagedorn,
Astron. Astrophys. {\bf 5}, 184 (1970).
\bibitem{hawking1974}
S.~W.~Hawking, Nature {\bf 248}, 30 (1974).
\bibitem{hawking1975}
S.~W.~Hawking, Comm. Math. Phys. {\bf 43}, 199 (1975).
\bibitem{bd1982}
N.~D.~Birrel and P.~C.~W.~Davies,
{\it Quantum fields in curved space}, (Cambridge University Press,
	Cambridge, 1982).
\bibitem{nj1998}
J.~C.~Niemeyer and K.~Jedamzik,
Phys. Rev. Lett. {\bf 80} 5481 (1998).
\bibitem{yokoyama1998}
J.~Yokoyama,
Phys. Rev. D{\bf 58} 107502 (1998).
\bibitem{gl1999}
A.~M.~Green and A.~R.~Liddle,
Phys. Rev. D{\bf 60} 063509 (1999).
\bibitem{chisholm2006}
J. R. Chisholm, Phys. Rev. D{\bf 74}, 043512 (2006).
\bibitem{penrose1979}
R.~Penrose, {\it General Relativity: an Einstein Centenary Survey}, 
ed S. W. Hawking and W. Israel (Cambridge University Press, Cambridge,
	1979).
\bibitem{mena_tavakol1999}
F.~C.~Mena and R.~Tavakol, Class. Quantum Grav. {\bf 16}, 435 (1999).
\end{thebibliography}
\end{document}